\documentclass[preprint2,longauthor,trackchanges]{aastex63}

\usepackage[nameinlink]{cleveref}
\usepackage{ragged2e}
\usepackage[]{natbib}
\newcommand\degree{$^{\circ}$}
\newcommand\degrees\degree

\newcommand\jwst{\em JWST}
\newcommand\hst{\em HST}

\newcommand\origins{\em Origins}

\newcommand\rsun{R$_{\odot}$}
\newcommand\rjup{R$_{J}$}

\DeclareSymbolFont{UPM}{U}{eur}{m}{n}
\DeclareMathSymbol{\umu}{0}{UPM}{"16}
\let\oldumu=\umu
\renewcommand\umu{\ifmmode\oldumu\else\math{\oldumu}\fi}

\newcommand\microns \micron

\let\oldsim=\sim
\renewcommand\sim{\ifmmode\oldsim\else\math{\oldsim}\fi}
\let\oldpm=\pm
\renewcommand\pm{\ifmmode\oldpm\else\math{\oldpm}\fi}
\newcommand\by{\ifmmode\times\else\math{\times}\fi}

\newbox{\wdbox}
\renewcommand\c{\setbox\wdbox=\hbox{,}\hspace{\wd\wdbox}}
\renewcommand\i{\setbox\wdbox=\hbox{i}\hspace{\wd\wdbox}}
\newcommand\n{\hspace{0.5em}}

\newcount\timect
\newcount\hourct
\newcount\minct
\newcommand\now{\timect=\time \divide\timect by 60
         \hourct=\timect \multiply\hourct by 60
         \minct=\time \advance\minct by -\hourct
         \number\timect:\ifnum \minct < 10 0\fi\number\minct}

\catcode`@=11

\newcommand\comment[1]{}
\newcommand\commenton{\catcode`\%=14}
\newcommand\commentoff{\catcode`\%=12}

\renewcommand\math[1]{$#1$}
\newcommand\mathshifton{\catcode`\$=3}
\newcommand\mathshiftoff{\catcode`\$=12}

\comment{the backslash is necessary}

\comment{alignment tab}

\let\atab=&
\newcommand\atabon{\catcode`\&=4}
\newcommand\ataboff{\catcode`\&=12}

\let\oldmsp=\sp
\let\oldmsb=\sb
\def\sp#1{\ifmmode
           \oldmsp{#1}%
         \else\strut\raise.85ex\hbox{\scriptsize #1}\fi}
\def\sb#1{\ifmmode
           \oldmsb{#1}%
         \else\strut\raise-.54ex\hbox{\scriptsize #1}\fi}
\newbox\@sp
\newbox\@sb
\def\sbp#1#2{\ifmmode%
           \oldmsb{#1}\oldmsp{#2}%
         \else
           \setbox\@sb=\hbox{\sb{#1}}%
           \setbox\@sp=\hbox{\sp{#2}}%
           \rlap{\copy\@sb}\copy\@sp
           \ifdim \wd\@sb >\wd\@sp
             \hskip -\wd\@sp \hskip \wd\@sb
           \fi
        \fi}
\def\msp#1{\ifmmode
           \oldmsp{#1}
         \else \math{\oldmsp{#1}}\fi}
\def\msb#1{\ifmmode
           \oldmsb{#1}
         \else \math{\oldmsb{#1}}\fi}
\def\supon{\catcode`\^=7}
\def\supoff{\catcode`\^=12}
\def\subon{\catcode`\_=8}
\def\suboff{\catcode`\_=12}
\def\supsubon{\supon \subon}
\def\supsuboff{\supoff \suboff}

\newcommand\actcharon{\catcode`\~=13}
\newcommand\actcharoff{\catcode`\~=12}

\newcommand\paramon{\catcode`\#=6}
\newcommand\paramoff{\catcode`\#=12}

\comment{And now to turn us totally on and off...}

\newcommand\reservedcharson{\commenton \mathshifton \atabon \supsubon \actcharon
	\paramon}

\newcommand\reservedcharsoff{\commentoff \mathshiftoff \ataboff
	\supsuboff \actcharoff \paramoff}

\catcode`@=12
\reservedcharsoff

\reservedcharson

\comment{ Must have ONLY ONE of these... trust these macros, they work

}

\newcommand{\squishlist}{
 \begin{list}{$\bullet$}
  { \setlength{\itemsep}{0pt}
     \setlength{\parsep}{0pt}
     \setlength{\topsep}{0pt}
     \setlength{\partopsep}{0pt}
     \setlength{\leftmargin}{2.0em}
     \setlength{\labelwidth}{1.5em}
     \setlength{\labelsep}{0.5em} } }

\newcommand{\squishlisttwo}{
 \begin{list}{$\bullet$}
  { \setlength{\itemsep}{1pt}
     \setlength{\parsep}{3pt}
     \setlength{\topsep}{3pt}
     \setlength{\partopsep}{0pt}
     \setlength{\leftmargin}{2.0em}
     \setlength{\labelwidth}{1.5em}
     \setlength{\labelsep}{0.5em} } }

\newcommand{\squishend}{
  \end{list}  }

\reservedcharson
\actcharon

\accepted{July 16, 2020}
\submitjournal{ApJL}

\shorttitle{ExoPIE}
\shortauthors{Stevenson}
\graphicspath{{./}{figures/}}
\begin{document}

\title{A New Method For Studying Exoplanet Atmospheres\\ Using Planetary Infrared Excess}

\correspondingauthor{Kevin Stevenson}
\email{Kevin.Stevenson@jhuapl.edu}

\author[0000-0002-7352-7941]{Kevin B. Stevenson}
\affiliation{Johns Hopkins APL, 11100 Johns Hopkins Rd, Laurel, MD 20723, USA}

\collaboration{1}{(Space Telescopes Advanced Research Group on the Atmospheres of Transiting Exoplanets)}




\begin{abstract}

To date, the ability for observers to reveal the composition or thermal structure of an exoplanet's atmosphere has rested on two techniques: high-contrast direct imaging and time-series observations of transiting exoplanets.  The former is currently limited to characterizing young, massive objects while the latter requires near 90{\degree} orbital inclinations, thus limiting atmospheric studies to a small fraction of the total exoplanet population.  Here we present an observational and analysis technique for studying the atmospheres of non-transiting exoplanets that relies on acquiring simultaneous, broad-wavelength spectra and resolving planetary infrared emission from the stellar spectrum.  This method could provide an efficient means to study exoplanet atmospheric dynamics using sparsely-sampled phase curve observations or a mechanism to search for signs of life on non-transiting exoplanets orbiting the nearest M-dwarf stars (such as Proxima Centauri).  If shown to be effective with {\em James Webb Space Telescope} ({\jwst}) observations, the method of measuring planetary infrared excess (PIE) would open up the large population of nearby, non-transiting exoplanets for atmospheric characterization.

\end{abstract}



\section{Introduction} 
\label{sec:intro}

Using available instruments, it is currently impossible to determine an unresolved planet's absolute flux without observing it pass behind its host star.  This is because the secondary eclipse provides a reference point in time for which there is no planetary emission.  Non-transiting exoplanets do not have access to this reference point and, thus, currently only yield a relative flux when observed.
Early works with {\em Spitzer Space Telescope} \citep{SPITZER} observations of the non-transiting exoplanet $\upsilon$ Andromedae b demonstrate the challenges of producing reliable results with sparsely-sampled photometric data \citep{Harrington2006,Crossfield2010}.
With relatively-broad wavelength coverage, {\jwst}'s spectroscopic instruments will be capable of inferring the absolute flux of hot, and even warm, exoplanets by way of a reference point in wavelength. 

\section{Conceptualization} 
\label{sec:concept}

\begin{figure}[t]
    \centering
    \includegraphics[width=0.99\linewidth]{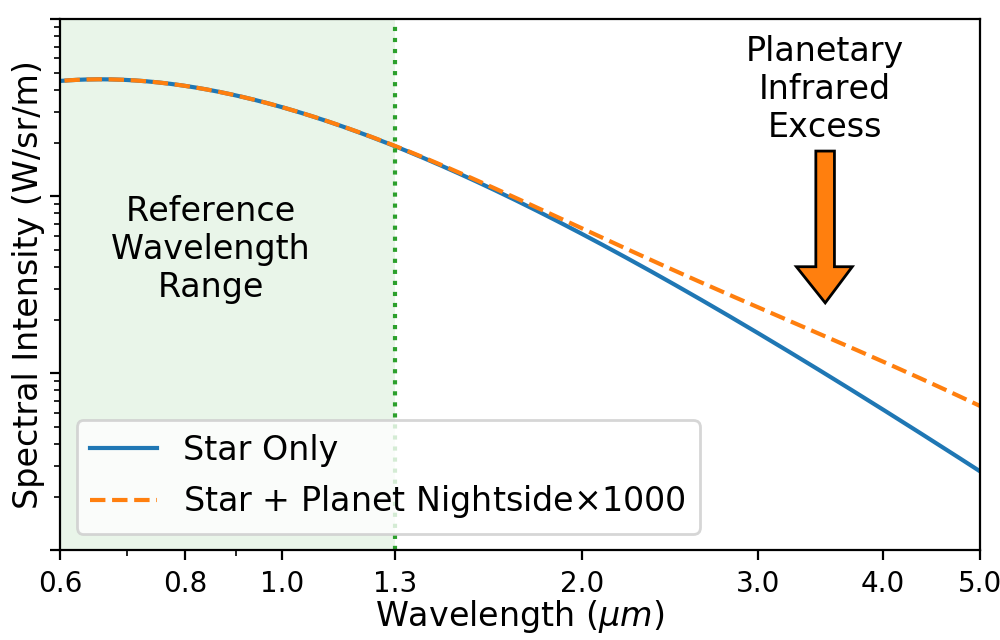}
    %\vspace*{-0.5\baselineskip}
    \caption{\label{fig:blackbody}{
    {\bf Toy figure depicting the concept of PIE.}  The spectral intensity of a 4,400~K blackbody (representing WASP-43) is shown with and without the addition of a 1,000~K blackbody (WASP-43b's nightside) that has been enhanced by a factor of 1,000 for visual clarity.  
    Conceptually, an atmospheric retrieval code can use spectroscopic data from the reference wavelength range ($<1.3$~{\microns} in this example) to constrain the stellar parameters and use the inferred infrared excess at longer wavelengths to determine the planetary parameters.  Spectroscopically resolving the sharp increase in planet flux on the Wien side of the blackbody is crucial to removing the planet's temperature/radius degeneracy.
    %temperature, radius, and/or atmospheric composition
    }}
\end{figure}

% Describe PIE
\Cref{fig:blackbody} provides a qualitative visualization of the planetary infrared excess (PIE) technique.  It depicts the spectral intensity from a 4,400~K blackbody with and without a 1,000~K blackbody that has been enhanced by a factor of 1,000.  Although this scenario applies to the exoplanet system WASP-43 with and without planetary nightside emission, it is also representative of many other hot Jupiters, which have been shown to have uniform nightside temperatures of $\sim$1,000~K \citep{Keating2019,Beatty2019}.  These warm planet nightsides emit no measurable flux at $<1.3$~{\microns}, thus providing a suitable reference wavelength range to constrain the stellar spectral energy distribution (SED).  When measured simultaneously at longer wavelengths ({\em e.g.,} $1.3-5$~{\microns}), one can compare the extrapolated stellar models to the observational data and infer that the measured infrared excess originates from the planet.  This is akin to performing a two-component SED fit to identify unresolved binary stars \citep{Burgasser2010,ElBadry2017} except here, the extreme difference in signal size between the two bodies requires the stability and precision of a space-based observatory. In general, the large temperature difference can help to distinguish the planet's peak SED in wavelength space from that of its host star.  The PIE method is also conceptually similar to attributing infrared excess to circumstellar disks when the stellar SED has a greater measured infrared flux than expected from that of a blackbody.
%If the system has a non-zero inclination (\textit{i.e.}, not aligned face on), the signal will often vary as a function of the planet's orbital phase, thus enabling observers to distinguish a planet from a circumstellar disk.  

\section{Demonstration} 
\label{sec:demo}

% W43 temperature constraint
\begin{figure*}[!t]
    \centering
    \includegraphics[width=0.49\linewidth]{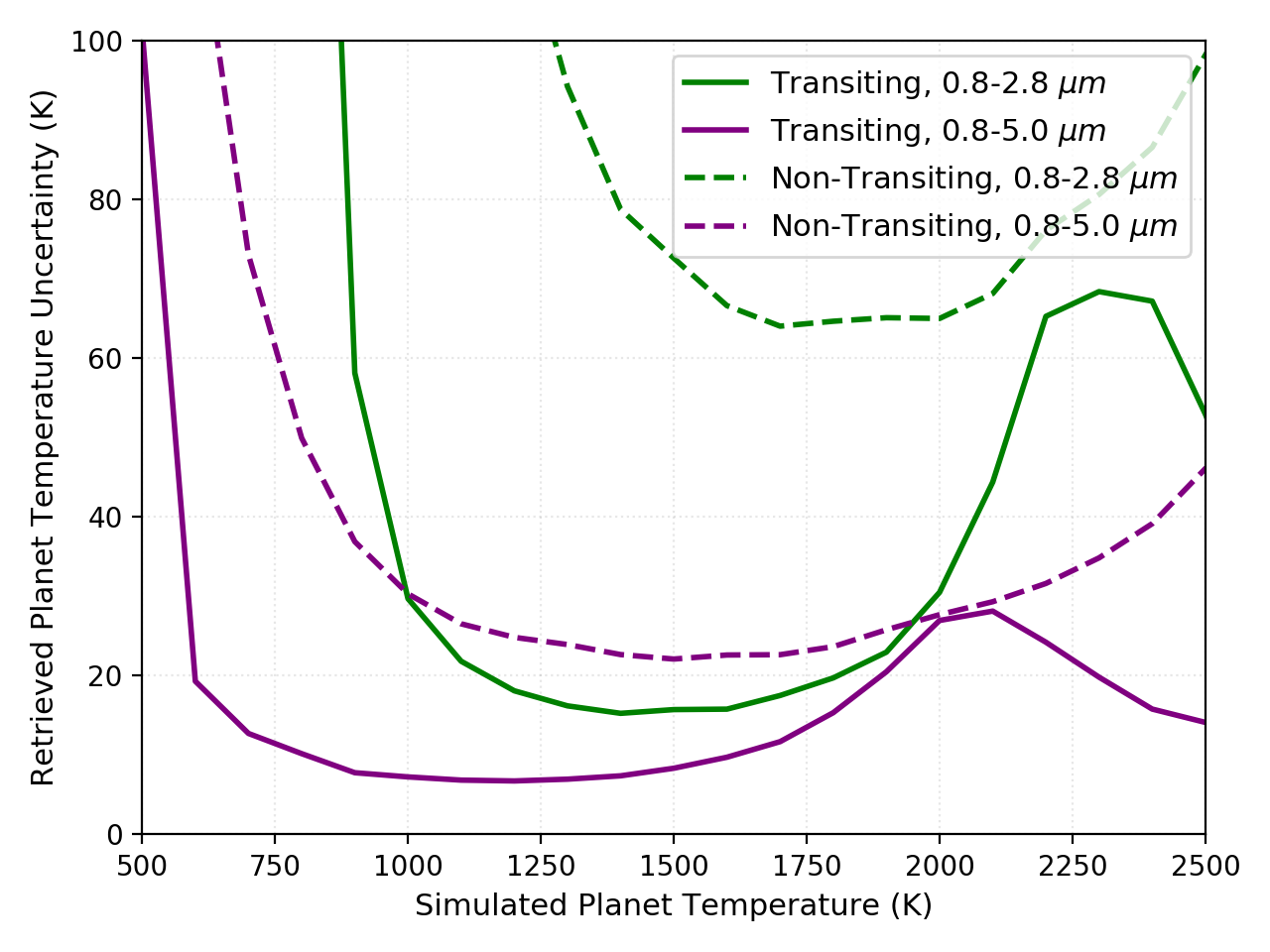}
    \includegraphics[width=0.49\linewidth]{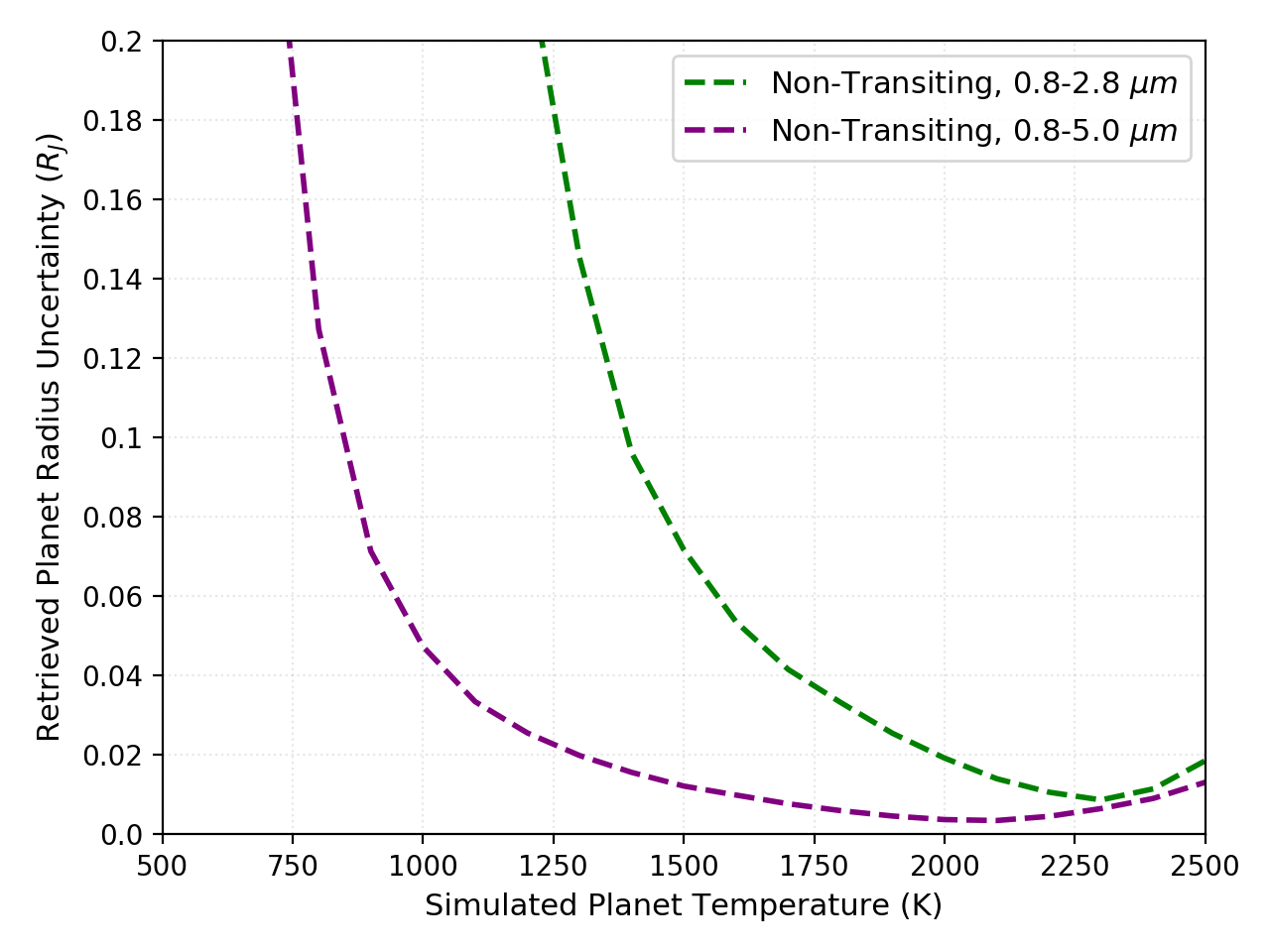}
    \vspace*{-0.5\baselineskip}
    \caption{\label{fig:TpRp}{
    {\bf Fitting stellar and planetary blackbody models simultaneously yields small uncertainties in retrieved planet temperatures ($\Delta T_P$, left) and radii ($\Delta R_P$, right) over a range of simulated planet nightside temperatures.}  Uncertainties are shown for both transiting (solid curves) and non-transiting (dashed curves) configurations, representing planets with known and unknown radii, respectively.  
    {\bf Left:} For our example exoplanetary system with blackbody emission (WASP-43, see \Cref{fig:blackbody}), the 1,000~K planet nightside is constrained to {$\Delta T_P = \pm$}30~K using one hour of {\jwst}'s NIRISS/SOSS mode (0.8--2.8~{\microns}).  Broader wavelength coverage yields even smaller uncertainties ({$\Delta T_P = \pm$}7~K over 0.8--5.0~{\microns}).  The temperature of a non-transiting twin of WASP-43b is constrained with only moderately larger uncertainties by also fitting for the planet radius ($\Delta T_P = \pm$30~K over 0.8--5.0~{\microns}).
    {\bf Right:} For the non-transiting hot Jupiters in our simulations, the planet radius is well constrained with NIRISS/SOSS alone ($\Delta R_P < 0.1$~{\rjup} for $T>1,400$~K and 0.8 -- 2.8~{\microns}). Adding wavelength coverage at longer wavelengths reduces the radius uncertainty appreciably for cooler planets ($\Delta R_P < 0.1$~{\rjup} for $T>850$~K and 0.8 -- 5.0~{\microns}).
    }}
\end{figure*}

As an initial demonstration of the PIE technique, we simulate WASP-43 ($R_S=0.667$~\rsun) as a $T_S=4,400$~K blackbody and WASP-43b ($R_P=1.036$~\rjup) over a range of nightside blackbody temperatures ($T_P=500-2,500$~K).  We consider two wavelength ranges representing {\jwst}'s NIRISS/SOSS instrument mode alone \citep[0.8 -- 2.8~\microns,][]{Doyon2012} and paired with {\jwst}'s NIRSpec/G395H mode \citep[0.8 -- 5.0~{\microns} total,][]{Ferruit2012}.  We use PandExo to generate realistic uncertainties \citep{Batalha2017}, assuming one hour of observation per mode.  When retrieving the best fit parameters ($R_S$, $T_S$, $R_P$, $T_P$), we adopt a fixed planet radius to simulate a transiting planet of known size and a uniform prior to simulate a non-transiting planet of unknown size.  We assume zero albedo for all cases; though a planet's nightside flux has no reflected light component.

\Cref{fig:TpRp} depicts the results of our experiment.  For each of our four simulations, we see a ``sweet spot'' in planet temperature that yields the smallest uncertainties (as precise as $\pm$7~K).  Increasing the wavelength coverage to 5.0 {\microns} extends the sweet spot to cooler planets.  Evidently, resolving the sharp increase in flux on the short-wavelength (Wien) side of the planet's blackbody emission is crucial to constraining its temperature, which is not possible with broadband photometric observations.
%yields significantly tighter temperature constraints for planets cooler than 1000~K.

When fitting for the planet radius, which is necessary for non-transiting exoplanets, there can be a strong degeneracy between the retrieved radius and temperature (see \Cref{fig:corner-jwst} in the Appendix).  This degeneracy leads to the larger temperature uncertainties seen in the left panel of \Cref{fig:TpRp}; however, utilizing broader wavelength coverage reduces its impact.  
The right panel of \Cref{fig:TpRp} illustrates the planet radius constraints from our two simulations with uniform priors.  Again, fitting for the planet radius benefits from utilizing broader wavelength coverage to resolve parameter degeneracies.
Stitching together spectra from multiple instrument modes (in this case NIRISS/SOSS and NIRSpec/G395H) will be explored early in Cycle 1 with {\jwst}'s Transiting Exoplanet Early Release Science Program \citep{Bean2018}.

\section{Applications} 
\label{sec:app}

\subsection{Sparsely-Sampled Phase Curves}

% Composition and phase curves
In addition to determining a planet's radius and brightness temperature, constraining the wavelength dependence of the infrared excess would provide information about its atmospheric composition.  Furthermore, this concept could be used to constrain the longitudinal variations in composition of an exoplanet atmosphere \citep{Stevenson2017a} using sparsely-sampled phase curve observations \citep{Krick2016}.
Currently, most exoplanet phase curves utilize time-consuming observations that start just prior to secondary eclipse and end shortly after the subsequent eclipse \citep[e.g.,][]{Knutson2012,Maxted2013,Mansfield2020}.  Having two reference points in time allows the observer to constrain the planet's nightside emission, which typically occurs near primary transit.  With the PIE technique, every spectroscopic frame provides a reference point in wavelength space that could, in principle, be calibrated against instrument or stellar variability \citep{Arcangeli2020,Wakeford2019}.  
Sparsely-sampled phase curves would be an ideal means to study atmospheric dynamics over a wide range of orbital periods and planet properties.
%and points to future successes with {\jwst}.

Even for the quietest and slowest rotators, stellar variability can cause changes to the measured spectrum that are orders of magnitude larger than a planet's emission signal.  Stellar variability could be removed by first reconstructing the measured stellar flux using multiple temperature-dependent components representing spots and/or plages on a constant photosphere.  \citet{Wakeford2019}  demonstrate this step by successfully reconstructing TRAPPIST-1's spectrum using three weighted stellar model components.  Next, to build up a more complete stellar model that spans the duration of an observation, the reconstructed stellar spectrum would need time-dependent components.  These can vary smoothly with stellar rotation or be stochastic events like flares.  Differentiating exceptionally cool plages from planetary emission should be feasible so long as the stellar rotational and planet's orbital periods are not related by a ratio of small integers.  Accurate stellar flare models will be critical toward fitting time-series data of active M-dwarf stars.  Finally, by evaluating the reconstructed stellar model on a frame-by-frame basis, one can build a reliable, long-time-baseline reference point in wavelength.  Dividing the time-series data by this model will yield the planet's phase-dependent emission spectrum.

\subsection{Nightside Temperature with Transit Observations}

%Use secondary eclipse to measure stellar spectra with PIE
Initial tests of the PIE technique with {\jwst} data should focus on transiting exoplanets with measured secondary eclipses.  For single-planet systems (like most hot Jupiter systems) data acquired during secondary eclipse will yield a pristine stellar spectrum.  Assuming any instrument or stellar variability is mostly achromatic, the measured stellar spectrum can be scaled to fit the reference wavelength range (see \Cref{fig:blackbody}) for individual frames acquired at any other orbital phase (such as primary transit or quadrature).  Recall that any discrepancy between the scaled stellar spectrum and the observed flux at longer wavelengths can be attributed to the planet.  Thus, baseline time-series data acquired outside of primary transit could be used to constrain a planet's nightside emission.  Having both a primary transit and secondary eclipse observation over the same wavelength range should be sufficient to measure a planet's day-night heat redistribution, thus potentially removing the need for full-orbit phase curve observations.
If shown to be effective, the PIE technique could open the door for {\jwst} to measure the atmospheric heat redistribution of any hot Jupiter exoplanet with both transit and eclipse observations.
%and not just those with the shortest orbital periods.
%with orbital periods greater than $\sim$2 days. 

Recent work using the {\em Hubble Space Telescope (HST)} to measure the relative thermal emission of WASP-12b at quadrature \citep{Arcangeli2020} demonstrates the real-world feasibility of the PIE technique when faced with relatively-large instrument or astrophysical systematics.  Their technique relies on applying common-mode corrections to remove instrument systematics then comparing the extracted spectra at each {\hst} orbit to the pristine stellar spectrum measured during eclipse.  They attribute any differences in flux to planetary emission.
WASP-12b's measured brightness temperature at quadrature (2,124{\pm}314~K) is consistent with {\em Spitzer Space Telescope} constraints from full-orbit phase curves at 3.6 and 4.5 {\microns}.  \citet{Arcangeli2020} were unable to extract the planet's nightside temperature due to a shift in the position of the spectrum on the detector relative to the eclipse and quadrature observations.

\subsection{Direct Imaging}

% Sun - Direct Imaging
\begin{figure}[!t]
    \centering
    \includegraphics[width=0.99\linewidth]{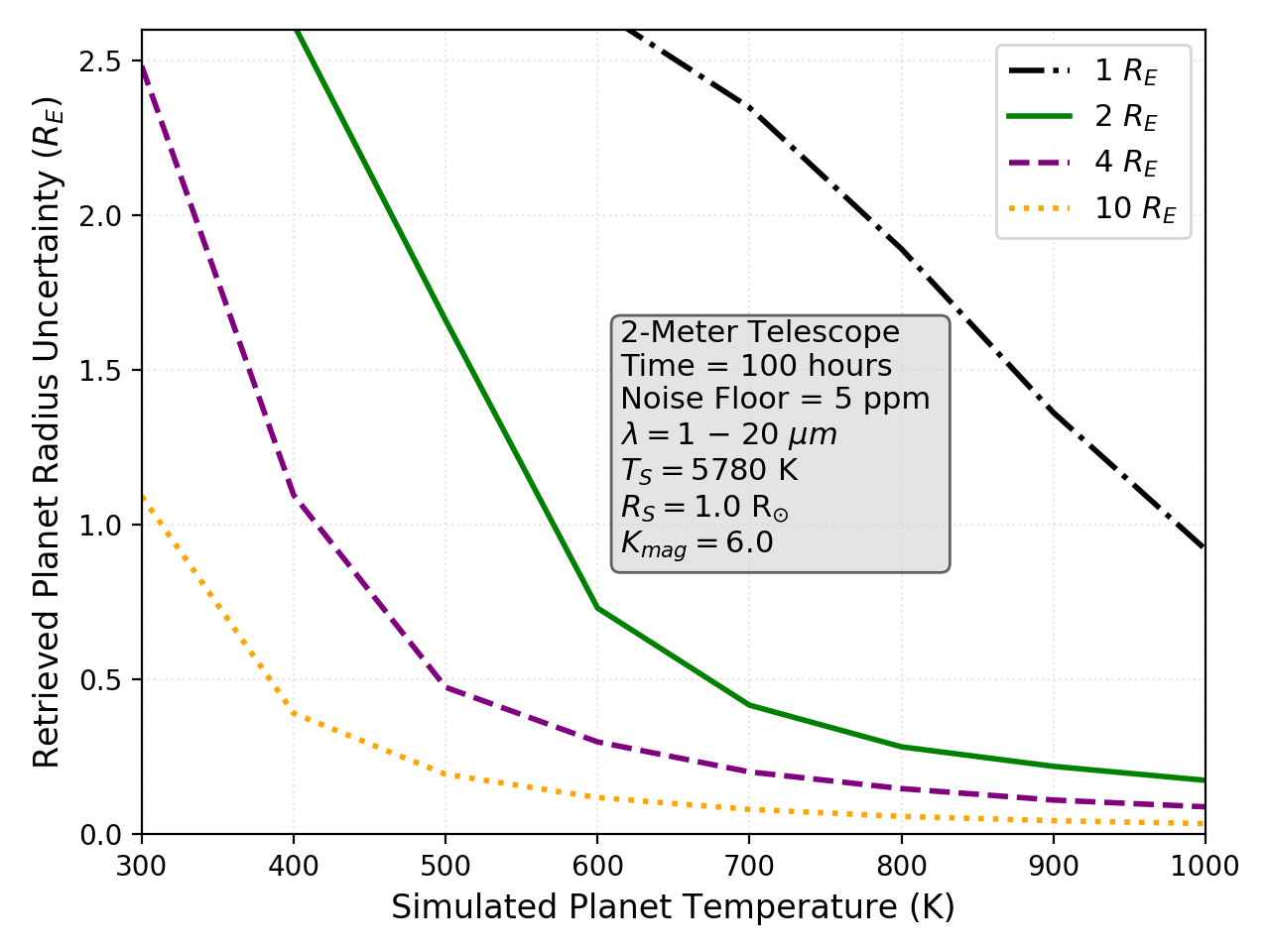}
    %\vspace*{-0.5\baselineskip}
    \caption{\label{fig:Sun}{
    {\bf A 2-meter class telescope with simultaneous, broad-wavelength coverage could constrain the radius of warm, hydrogen-dominated exoplanets orbiting Sun-like stars.}  We simulate a 2-meter version of the {\em Origins Space Telescope} with \mbox{1 -- 20} {\micron} coverage to constrain stellar and planetary parameters using 100 hours of observing time per simulation.  At 700~K, the radii of all simulated planets with expected hydrogen-dominated atmospheres (2 -- 10~$R_E$) are reasonably constrained (bounded uncertainty $< {\pm}0.5~R_E$).  By 400~K, only the size of the ten Earth-radii planet is reasonably constrained, while the size of the two Earth-radii planet is defined by an upper limit.
    %The size of a two Earth-radii planet is reasonably constrained (uncertainty $< {\pm}0.5~R_E$) at $>700$~K, whereas a ten Earth-radii planet is reasonably constrained at $>400$~K.
    }}
\end{figure}

%Direct imaging radius
In addition to simulating Jupiter-size exoplanets, we investigate how the PIE technique could be used to constrain the radii of long-period planets discovered by future high-contrast direct imaging missions such as the {\em Roman Space Telescope}, HabEx, or LUVOIR \citep{Akeson2019,HabEx2020,LUVOIR2019}.
%First, planets discovered using high-contrast direct imaging often have unconstrained radii.  
Being able to differentiate between Earth- and Neptune-size planets is particularly important in the search for signs of life on rocky worlds.
For planets around Sun-like stars, \Cref{fig:Sun} demonstrates that retrievals using the PIE technique can constrain the size of warm Neptunes ($T_P > 500$~K, $R_P \geq 4 R_E$) to better than ${\pm}0.5 R_E$, but would be unable to place meaningful radius constraints on potentially-habitable worlds.
%place meaningful upper limits on temperature and radius for these directly imaged planets.  
%Requires 1-20 microns to constrain both stellar and planet parameters

\subsection{Habitability}

%Origins
Finally, we consider how the PIE technique could be used to help future missions search for signs of life. % on the nearest, habitable zone planets.  
The {\em Origins Space Telescope} (hereafter {\origins}) mission concept was designed to search for biosignatures on planets transiting the nearest M-dwarf stars \citep{OST2019}.  Relative to Sun-like stars, M-dwarf planet-to-star flux ratios are much more favorable and, thus, the PIE technique should yield more precise planet temperatures and radii. %even for terrestrial worlds.
The first two steps in {\origins}' exoplanet observing strategy are to confirm the presence of an atmosphere and constrain the planet's surface temperature to be within the habitable zone.
For non-transiting exoplanets, the first step could be readily accomplished by comparing the measured dayside and nightside emission.  Assuming non-zero orbital inclinations, planets with tenuous atmospheres (like Mars) will exhibit strong day-night contrasts, whereas those with thick atmospheres (like Venus) will display no measurable variation \citep{Seager2009,Selsis2011,Kreidberg2019}.  Importantly, it may be possible to constrain these planets' orbital inclinations with multi-epoch, high-resolution, cross-correlation techniques \citep{Buzard2020}.
% Prox Cen
\begin{figure}[!t]
    \centering
    \includegraphics[width=0.99\linewidth]{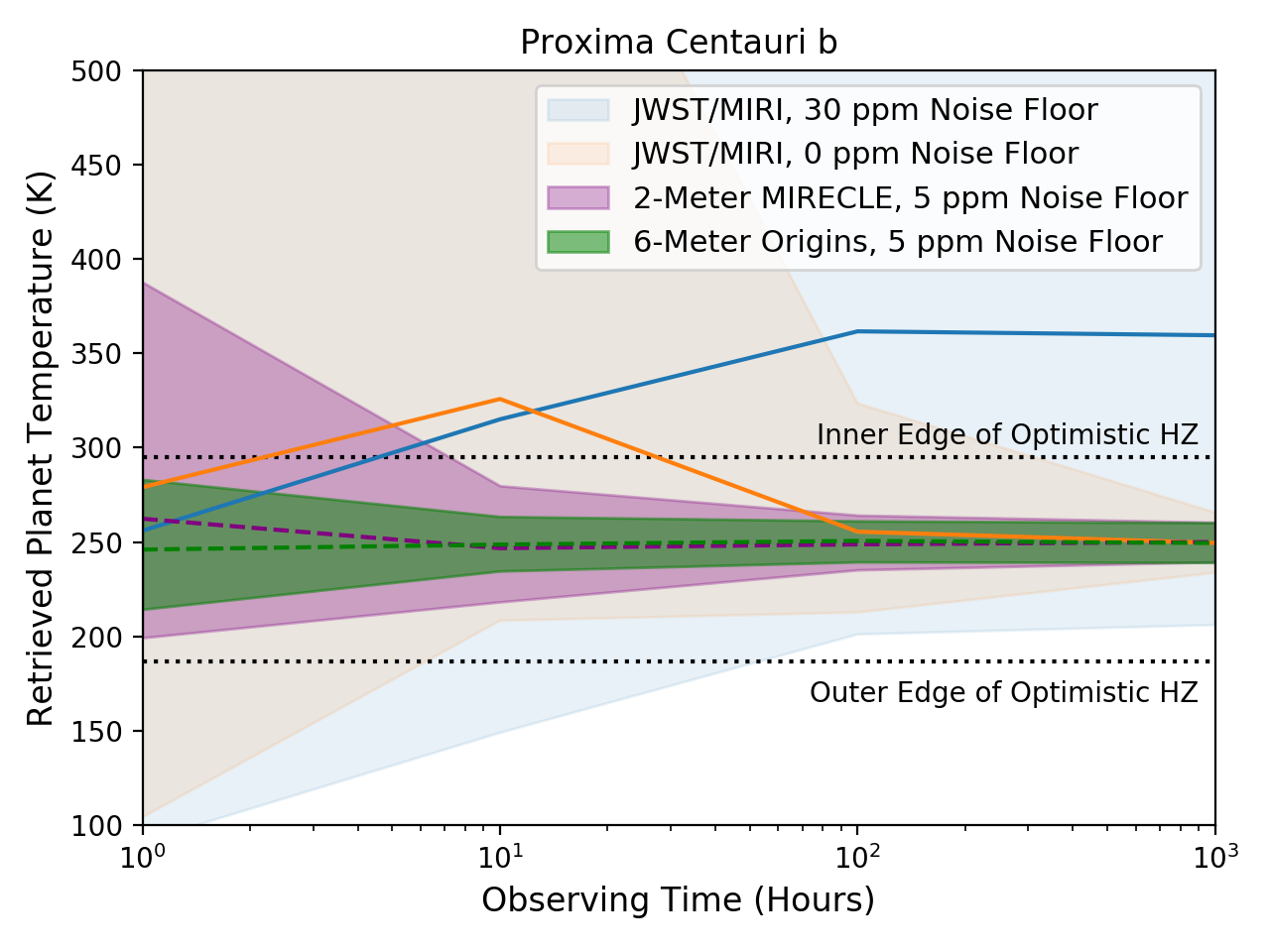}
    %\vspace*{-0.5\baselineskip}
    \caption{\label{fig:ProxCen}{
    {\bf With their simultaneous, broad-wavelength coverage, the {\origins} and {\em MIRECLE} mission concepts could identify which nearby M-dwarf planets can sustain liquid water on their surfaces.}  We apply the PIE technique to a simulation of the nearest habitable zone (HZ) exoplanet, Proxima Centauri b, using four telescope/instrument configurations, {\jwst}/MIRI (5 -- 13 {\microns}, 0 and 30 ppm noise floor), {\em MIRECLE} (3 -- 20 {\microns}, 2-meter aperture), and {\origins} (3 -- 20 {\microns}, 6-meter aperture).  {\jwst} is unlikely to precisely constrain the planet's temperature without significant telescope time ($>$1,000 hours) and a much lower noise floor than current expectations ($\ll$30~ppm).  With 10 hours of integration time, {\em Origins} can constrain the simulated planet's temperature (green shaded region) to within {$\pm$}15~K and its radius to within {$\pm$}0.1 Earth radii (not shown).  Being a smaller version of {\origins}, {\em MIRECLE} requires $\sim$100 hours to achieve the same precision (purple shaded region).  Lines depict the median of the retrieved temperatures and shaded regions depict $1\sigma$ uncertainties.
    }}
\end{figure}

We demonstrate how the PIE technique could achieve {\origins}' second step in their observing strategy by simulating \mbox{Proxima Centauri b}, the nearest, potentially-habitable exoplanet \citep{Anglada2016}.  \Cref{fig:ProxCen} illustrates the retrieved planet temperature (with 1$\sigma$ uncertainties) as a function of observing time for three telescopes: {\jwst}, {\origins}, and {\em MIRECLE} \citep[Mid-InfraRed Exoplanet CLimate Explorer,][]{Staguhn2019}.  Regardless of its estimated noise floor, {\jwst}'s MIRI/LRS mode \citep[Mid-Infrared Instrument / Low Resolution Spectrometer, 5 -- 13 {\microns};][]{MIRI2018} is unlikely to precisely constrain Proxima Centauri b's temperature or radius due to having insufficient coverage at longer wavelengths.  
With its broad, simultaneous wavelength coverage (3 -- 20 {\microns}), the 6-meter {\origins} is predicted to constrain Proxima Centauri b's brightness temperature to be within the inner and outer edges of the optimistic habitable zone \citep{Kopparapu2017} with confidences of 3.5 and 5.7$\sigma$, respectively, by observing the system for 10 hours.  A 2-meter {\em MIRECLE} mission concept (also 3 -- 20 {\microns}) would require $\sim$100 hours to achieve the same detection significance.

The third (and final) step in {\origins}'s observing strategy is to search for biosignatures.  There are numerous molecular features in the mid-infrared that would be good indicators of life ({\em e.g.}, O$_3$+CH$_4$ or O$_3$+N$_2$O); however, more work is needed to determine whether or not the PIE technique could be used to identify these pairs of molecular features in non-transiting exoplanet atmospheres (see the Appendix).  If shown to be capable, a mid-infrared mission (such as {\origins}) could characterize the atmospheres of $\sim$240 temperate, terrestrial worlds within 15~parsec \citep{Barclay2018}.  Many of these non-transiting exoplanets will be discovered this decade by ground-based radial velocity surveys focused on identifying planets orbiting the nearest M-dwarf stars \citep{Crepp2016,Claudi2018,Seifahrt2018}.

\software{PandExo \citep{Batalha2017},
          DYNESTY \citep{Speagle2020},
          EAOT \citep{Mullally2019}}

\acknowledgments

We appreciate the feedback of N. Lewis, J. Bean, and J.-M. D{\'e}sert upon reviewing early versions of the manuscript.  This work was supported through the James Webb Space Telescope Early Release Science program (JWST-ERS-01366) and internal funding at Johns Hopkins Applied Physics Laboratory.

\bibliography{ms}

\begin{thebibliography}{}
\expandafter\ifx\csname natexlab\endcsname\relax\def\natexlab#1{#1}\fi

\bibitem[{Akeson {et~al.}(2019)Akeson, Armus, Bachelet, Bailey, Bartusek,
  Bellini, Benford, Bennett, Bhattacharya, Bohlin, Boyer, Bozza, Bryden,
  Novati, Carpenter, Casertano, Choi, Content, Dayal, Dressler, Doré, Fall,
  Fan, Fang, Filippenko, Finkelstein, Foley, Furlanetto, Kalirai, Gaudi,
  Gilbert, Girard, Grady, Greene, Guhathakurta, Heinrich, Hemmati, Hendel,
  Henderson, Henning, Hirata, Ho, Huff, Hutter, Jansen, Jha, Johnson, Jones,
  Kasdin, Kelly, Kirshner, Koekemoer, Kruk, Lewis, Macintosh, Madau, Malhotra,
  Mandel, Massara, Masters, McEnery, McQuinn, Melchior, Melton, Mennesson,
  Peeples, Penny, Perlmutter, Pisani, Plazas, Poleski, Postman, Ranc, Rauscher,
  Rest, Roberge, Robertson, Rodney, Rhoads, Rhodes, Jr., Sahu, Sand, Scolnic,
  Seth, Shvartzvald, Siellez, Smith, Spergel, Stassun, Street, Strolger,
  Szalay, Trauger, Troxel, Turnbull, van~der Marel, von~der Linden, Wang,
  Weinberg, Williams, Windhorst, Wollack, Wu, Yee, \& Zimmerman}]{Akeson2019}
Akeson, R., Armus, L., Bachelet, E., {et~al.} 2019, The Wide Field Infrared
  Survey Telescope: 100 Hubbles for the 2020s, arXiv:1902.05569

\bibitem[{{Anglada-Escud{\'e}} {et~al.}(2016){Anglada-Escud{\'e}}, {Amado},
  {Barnes}, {Berdi{\~n}as}, {Butler}, {Coleman}, {de La Cueva}, {Dreizler},
  {Endl}, {Giesers}, {Jeffers}, {Jenkins}, {Jones}, {Kiraga}, {K{\"u}rster},
  {L{\'o}pez-Gonz{\'a}lez}, {Marvin}, {Morales}, {Morin}, {Nelson}, {Ortiz},
  {Ofir}, {Paardekooper}, {Reiners}, {Rodr{\'\i}guez},
  {Rodr{\'\i}guez-L{\'o}pez}, {Sarmiento}, {Strachan}, {Tsapras}, {Tuomi}, \&
  {Zechmeister}}]{Anglada2016}
{Anglada-Escud{\'e}}, G., {Amado}, P.~J., {Barnes}, J., {et~al.} 2016, \nat,
  536, 437

\bibitem[{{Arcangeli} {et~al.}(2020){Arcangeli}, {Desert}, {Parmentier},
  {Tsai}, \& {Stevenson}}]{Arcangeli2020}
{Arcangeli}, J., {Desert}, J.-M., {Parmentier}, V., {Tsai}, S.-M., \&
  {Stevenson}, K.~B. 2020, Submitted

\bibitem[{{Barclay} {et~al.}(2018){Barclay}, {Pepper}, \&
  {Quintana}}]{Barclay2018}
{Barclay}, T., {Pepper}, J., \& {Quintana}, E.~V. 2018, apjs, 239, 2

\bibitem[{{Batalha} {et~al.}(2017){Batalha}, {Mandell}, {Pontoppidan},
  {Stevenson}, {Lewis}, {Kalirai}, {Earl}, {Greene}, {Albert}, \&
  {Nielsen}}]{Batalha2017}
{Batalha}, N.~E., {Mandell}, A., {Pontoppidan}, K., {et~al.} 2017, \pasp, 129,
  064501

\bibitem[{{Bean} {et~al.}(2018){Bean}, {Stevenson}, {Batalha},
  {Berta-Thompson}, {Kreidberg}, {Crouzet}, {Benneke}, {Line}, {Sing},
  {Wakeford}, {Knutson}, {Kempton}, {D{\'e}sert}, {Crossfield}, {Batalha}, {de
  Wit}, {Parmentier}, {Harrington}, {Moses}, {Lopez-Morales}, {Alam}, {Blecic},
  {Bruno}, {Carter}, {Chapman}, {Decin}, {Dragomir}, {Evans}, {Fortney},
  {Fraine}, {Gao}, {Garc{\'\i}a Mu{\~n}oz}, {Gibson}, {Goyal}, {Heng}, {Hu},
  {Kendrew}, {Kilpatrick}, {Krick}, {Lagage}, {Lendl}, {Louden}, {Madhusudhan},
  {Mandell}, {Mansfield}, {May}, {Morello}, {Morley}, {Nikolov}, {Redfield},
  {Roberts}, {Schlawin}, {Spake}, {Todorov}, {Tsiaras}, {Venot}, {Waalkes},
  {Wheatley}, {Zellem}, {Angerhausen}, {Barrado}, {Carone}, {Casewell},
  {Cubillos}, {Damiano}, {de Val-Borro}, {Drummond}, {Edwards}, {Endl},
  {Espinoza}, {France}, {Gizis}, {Greene}, {Henning}, {Hong}, {Ingalls}, {Iro},
  {Irwin}, {Kataria}, {Lahuis}, {Leconte}, {Lillo-Box}, {Lines}, {Lothringer},
  {Mancini}, {Marchis}, {Mayne}, {Palle}, {Rauscher}, {Roudier}, {Shkolnik},
  {Southworth}, {Swain}, {Taylor}, {Teske}, {Tinetti}, {Tremblin}, {Tucker},
  {van Boekel}, {Waldmann}, {Weaver}, \& {Zingales}}]{Bean2018}
{Bean}, J.~L., {Stevenson}, K.~B., {Batalha}, N.~M., {et~al.} 2018, PASP, 130,
  114402

\bibitem[{{Beatty} {et~al.}(2019){Beatty}, {Marley}, {Gaudi}, {Col{\'o}n},
  {Fortney}, \& {Showman}}]{Beatty2019}
{Beatty}, T.~G., {Marley}, M.~S., {Gaudi}, B.~S., {et~al.} 2019, aj, 158, 166

\bibitem[{Burgasser {et~al.}(2010)Burgasser, Cruz, Cushing, Gelino, Looper,
  Faherty, Kirkpatrick, \& Reid}]{Burgasser2010}
Burgasser, A.~J., Cruz, K.~L., Cushing, M., {et~al.} 2010, The Astrophysical
  Journal, 710, 1142–1169

\bibitem[{{Buzard} {et~al.}(2020){Buzard}, {Finnerty}, {Piskorz}, {Pelletier},
  {Benneke}, {Bender}, {Lockwood}, {Wallack}, {Wilkins}, \&
  {Blake}}]{Buzard2020}
{Buzard}, C., {Finnerty}, L., {Piskorz}, D., {et~al.} 2020, \aj, 160, 1

\bibitem[{{Claudi} {et~al.}(2018){Claudi}, {Benatti}, {Carleo}, {Ghedina},
  {Guerra}, {Ghinassi}, {Harutyunyan}, {Micela}, {Molinari}, {Oliva}, {Rainer},
  {Tozzi}, {Baffa}, {Baruffolo}, {Biliotti}, {Buchschacher}, {Cecconi},
  {Cosentino}, {Falcini}, {Fantinel}, {Fini}, {Giani}, {Gonzalez-Alvarez},
  {Gonzalez}, {Gonzalez}, {Gratton}, {Hernandez}, {Iuzzolino}, {Lodi},
  {Malavolta}, {Maldonado}, {Origlia}, {Puglisi}, {Sanna}, {San Juan
  G{\'o}mez}, {Scuderi}, {Seemann}, {Sozzetti}, {Sozzi}, {Perez Ventura},
  {Hernandez Diaz}, {Galli}, {Riverol}, \& {Riverol}}]{Claudi2018}
{Claudi}, R., {Benatti}, S., {Carleo}, I., {et~al.} 2018, in Society of
  Photo-Optical Instrumentation Engineers (SPIE) Conference Series, Vol. 10702,
  Proc.~SPIE, 107020Z

\bibitem[{{Crepp} {et~al.}(2016){Crepp}, {Crass}, {King}, {Bechter}, {Bechter},
  {Ketterer}, {Reynolds}, {Hinz}, {Kopon}, {Cavalieri}, {Fantano}, {Koca},
  {Onuma}, {Stapelfeldt}, {Thomes}, {Wall}, {Macenka}, {McGuire}, {Korniski},
  {Zugby}, {Eisner}, {Gaudi}, {Hearty}, {Kratter}, {Kuchner}, {Micela},
  {Nelson}, {Pagano}, {Quirrenbach}, {Schwab}, {Skrutskie}, {Sozzetti},
  {Woodward}, \& {Zhao}}]{Crepp2016}
{Crepp}, J.~R., {Crass}, J., {King}, D., {et~al.} 2016, in Society of
  Photo-Optical Instrumentation Engineers (SPIE) Conference Series, Vol. 9908,
  Proc.~SPIE, 990819

\bibitem[{{Crossfield} {et~al.}(2010){Crossfield}, {Hansen}, {Harrington},
  {Cho}, {Deming}, {Menou}, \& {Seager}}]{Crossfield2010}
{Crossfield}, I. J.~M., {Hansen}, B. M.~S., {Harrington}, J., {et~al.} 2010,
  apj, 723, 1436

\bibitem[{{Doyon} {et~al.}(2012){Doyon}, {Hutchings}, {Beaulieu}, {Albert},
  {Lafreni{\`e}re}, {Willott}, {Touahri}, {Rowlands}, {Maszkiewicz},
  {Fullerton}, {Volk}, {Martel}, {Chayer}, {Sivaramakrishnan}, {Abraham},
  {Ferrarese}, {Jayawardhana}, {Johnstone}, {Meyer}, {Pipher}, \&
  {Sawicki}}]{Doyon2012}
{Doyon}, R., {Hutchings}, J.~B., {Beaulieu}, M., {et~al.} 2012, in Proc.~SPIE,
  Vol. 8442, Space Telescopes and Instrumentation 2012: Optical, Infrared, and
  Millimeter Wave, 84422R

\bibitem[{El-Badry {et~al.}(2017)El-Badry, Rix, Ting, Weisz, Bergemann,
  Cargile, Conroy, \& Eilers}]{ElBadry2017}
El-Badry, K., Rix, H.-W., Ting, Y.-S., {et~al.} 2017, Monthly Notices of the
  Royal Astronomical Society, 473, 5043

\bibitem[{{Feroz} {et~al.}(2009){Feroz}, {Hobson}, \& {Bridges}}]{Feroz2009}
{Feroz}, F., {Hobson}, M.~P., \& {Bridges}, M. 2009, MNRAS, 398, 1601

\bibitem[{{Ferruit} {et~al.}(2012){Ferruit}, {Bagnasco}, {Barho}, {Birkmann},
  {B{\"o}ker}, {De Marchi}, {Dorner}, {Ehrenwinkler}, {Falcolini}, {Giardino},
  {Gnata}, {Honnen}, {Jakobsen}, {Jensen}, {Kolm}, {Maier}, {Maurer}, {Melf},
  {Mosner}, {Rumler}, {Salvignol}, {Sirianni}, {Strada}, {te Plate}, \&
  {Wettemann}}]{Ferruit2012}
{Ferruit}, P., {Bagnasco}, G., {Barho}, R., {et~al.} 2012, in Society of
  Photo-Optical Instrumentation Engineers (SPIE) Conference Series, Vol. 8442,
  Society of Photo-Optical Instrumentation Engineers (SPIE) Conference Series,
  84422O

\bibitem[{{Gaudi} {et~al.}(2020){Gaudi}, {Seager}, {Mennesson}, {Kiessling},
  {Warfield}, {Cahoy}, {Clarke}, {Domagal-Goldman}, {Feinberg}, {Guyon},
  {Kasdin}, {Mawet}, {Plavchan}, {Robinson}, {Rogers}, {Scowen}, {Somerville},
  {Stapelfeldt}, {Stark}, {Stern}, {Turnbull}, {Amini}, {Kuan}, {Martin},
  {Morgan}, {Redding}, {Stahl}, {Webb}, {Alvarez-Salazar}, {Arnold}, {Arya},
  {Balasubramanian}, {Baysinger}, {Bell}, {Below}, {Benson}, {Blais}, {Booth},
  {Bourgeois}, {Bradford}, {Brewer}, {Brooks}, {Cady}, {Caldwell}, {Calvet},
  {Carr}, {Chan}, {Cormarkovic}, {Coste}, {Cox}, {Danner}, {Davis}, {Dewell},
  {Dorsett}, {Dunn}, {East}, {Effinger}, {Eng}, {Freebury}, {Garcia}, {Gaskin},
  {Greene}, {Hennessy}, {Hilgemann}, {Hood}, {Holota}, {Howe}, {Huang}, {Hull},
  {Hunt}, {Hurd}, {Johnson}, {Kissil}, {Knight}, {Kolenz}, {Kraus}, {Krist},
  {Li}, {Lisman}, {Mandic}, {Mann}, {Marchen}, {Marrese-Reading}, {McCready},
  {McGown}, {Missun}, {Miyaguchi}, {Moore}, {Nemati}, {Nikzad}, {Nissen},
  {Novicki}, {Perrine}, {Pineda}, {Polanco}, {Putnam}, {Qureshi}, {Richards},
  {Eldorado Riggs}, {Rodgers}, {Rud}, {Saini}, {Scalisi}, {Scharf}, {Schulz},
  {Serabyn}, {Sigrist}, {Sikkia}, {Singleton}, {Shaklan}, {Smith}, {Southerd},
  {Stahl}, {Steeves}, {Sturges}, {Sullivan}, {Tang}, {Taras}, {Tesch},
  {Therrell}, {Tseng}, {Valente}, {Van Buren}, {Villalvazo}, {Warwick}, {Webb},
  {Westerhoff}, {Wofford}, {Wu}, {Woo}, {Wood}, {Ziemer}, {Arney}, {Anderson},
  {Ma{\'\i}z-Apell{\'a}niz}, {Bartlett}, {Belikov}, {Bendek}, {Cenko},
  {Douglas}, {Dulz}, {Evans}, {Faramaz}, {Feng}, {Ferguson}, {Follette},
  {Ford}, {Garc{\'\i}a}, {Geha}, {Gelino}, {G{\"o}tberg}, {Hildebrand t}, {Hu},
  {Jahnke}, {Kennedy}, {Kreidberg}, {Isella}, {Lopez}, {Marchis}, {Macri},
  {Marley}, {Matzko}, {Mazoyer}, {McCandliss}, {Meshkat}, {Mordasini},
  {Morris}, {Nielsen}, {Newman}, {Petigura}, {Postman}, {Reines}, {Roberge},
  {Roederer}, {Ruane}, {Schwieterman}, {Sirbu}, {Spalding}, {Teplitz},
  {Tumlinson}, {Turner}, {Werk}, {Wofford}, {Wyatt}, {Young}, \&
  {Zellem}}]{HabEx2020}
{Gaudi}, B.~S., {Seager}, S., {Mennesson}, B., {et~al.} 2020, arXiv e-prints,
  arXiv:2001.06683

\bibitem[{{Harrington} {et~al.}(2006){Harrington}, {Hansen}, {Luszcz},
  {Seager}, {Deming}, {Menou}, {Cho}, \& {Richardson}}]{Harrington2006}
{Harrington}, J., {Hansen}, B.~M., {Luszcz}, S.~H., {et~al.} 2006, Science,
  314, 623

\bibitem[{{Keating} {et~al.}(2019){Keating}, {Cowan}, \& {Dang}}]{Keating2019}
{Keating}, D., {Cowan}, N.~B., \& {Dang}, L. 2019, Nature Astronomy, 3, 1092

\bibitem[{{Kendrew} {et~al.}(2018){Kendrew}, {Dicken}, {Bouwman}, {Garcia
  Marin}, {Greene}, {Lagage}, {Ressler}, {Crouzet}, {Kreidberg}, {Batalha},
  {Bean}, {Stevenson}, {Glasse}, {Wright}, \& {Rieke}}]{MIRI2018}
{Kendrew}, S., {Dicken}, D., {Bouwman}, J., {et~al.} 2018, in Society of
  Photo-Optical Instrumentation Engineers (SPIE) Conference Series, Vol. 10698,
  Proc.~SPIE, 106983U

\bibitem[{{Knutson} {et~al.}(2012){Knutson}, {Lewis}, {Fortney}, {Burrows},
  {Showman}, {Cowan}, {Agol}, {Aigrain}, {Charbonneau}, {Deming}, {D{\'e}sert},
  {Henry}, {Langton}, \& {Laughlin}}]{Knutson2012}
{Knutson}, H.~A., {Lewis}, N., {Fortney}, J.~J., {et~al.} 2012, apj, 754, 22

\bibitem[{{Kopparapu} {et~al.}(2017){Kopparapu}, {Wolf}, {Arney}, {Batalha},
  {Haqq-Misra}, {Grimm}, \& {Heng}}]{Kopparapu2017}
{Kopparapu}, R.~k., {Wolf}, E.~T., {Arney}, G., {et~al.} 2017, apj, 845, 5

\bibitem[{{Kreidberg} {et~al.}(2019){Kreidberg}, {Koll}, {Morley}, {Hu},
  {Schaefer}, {Deming}, {Stevenson}, {Dittmann}, {Vanderburg}, {Berardo},
  {Guo}, {Stassun}, {Crossfield}, {Charbonneau}, {Latham}, {Loeb}, {Ricker},
  {Seager}, \& {Vand erspek}}]{Kreidberg2019}
{Kreidberg}, L., {Koll}, D. D.~B., {Morley}, C., {et~al.} 2019, Nature, 573, 87

\bibitem[{{Krick} {et~al.}(2016){Krick}, {Ingalls}, {Carey}, {von Braun},
  {Kane}, {Ciardi}, {Plavchan}, {Wong}, \& {Lowrance}}]{Krick2016}
{Krick}, J.~E., {Ingalls}, J., {Carey}, S., {et~al.} 2016, \apj, 824, 27

\bibitem[{{Mansfield} {et~al.}(2020){Mansfield}, {Bean}, {Stevenson},
  {Komacek}, {Bell}, {Tan}, {Malik}, {Beatty}, {Wong}, {Cowan}, {Dang},
  {D{\'e}sert}, {Fortney}, {Gaudi}, {Keating}, {Kempton}, {Kreidberg}, {Line},
  {Parmentier}, {Stassun}, {Swain}, \& {Zellem}}]{Mansfield2020}
{Mansfield}, M., {Bean}, J.~L., {Stevenson}, K.~B., {et~al.} 2020, apjl, 888,
  L15

\bibitem[{{Maxted} {et~al.}(2013){Maxted}, {Anderson}, {Doyle}, {Gillon},
  {Harrington}, {Iro}, {Jehin}, {Lafreni{\`e}re}, {Smalley}, \&
  {Southworth}}]{Maxted2013}
{Maxted}, P.~F.~L., {Anderson}, D.~R., {Doyle}, A.~P., {et~al.} 2013, MNRAS,
  428, 2645

\bibitem[{{Meixner} {et~al.}(2019){Meixner}, {Cooray}, {Leisawitz}, {Staguhn},
  {Armus}, {Battersby}, {Bauer}, {Bergin}, {Bradford}, {Ennico-Smith},
  {Fortney}, {Kataria}, {Melnick}, {Milam}, {Narayanan}, {Padgett},
  {Pontoppidan}, {Pope}, {Roellig}, {Sandstrom}, {Stevenson}, {Su}, {Vieira},
  {Wright}, {Zmuidzinas}, {Sheth}, {Benford}, {Mamajek}, {Neff}, {De Beck},
  {Gerin}, {Helmich}, {Sakon}, {Scott}, {Vavrek}, {Wiedner}, {Carey},
  {Burgarella}, {Moseley}, {Amatucci}, {Carter}, {DiPirro}, {Wu}, {Beaman},
  {Beltran}, {Bolognese}, {Bradley}, {Corsetti}, {D'Asto}, {Denis}, {Derkacz},
  {Earle}, {Fantano}, {Folta}, {Gavares}, {Generie}, {Hilliard}, {Howard},
  {Jamil}, {Jamison}, {Lynch}, {Martins}, {Petro}, {Ramspacher}, {Rao}, {Sand
  in}, {Stoneking}, {Tompkins}, \& {Webster}}]{OST2019}
{Meixner}, M., {Cooray}, A., {Leisawitz}, D., {et~al.} 2019, arXiv e-prints,
  arXiv:1912.06213

\bibitem[{{Mullally} {et~al.}(2019){Mullally}, {Rodriguez}, {Stevenson}, \&
  {Wakeford}}]{Mullally2019}
{Mullally}, S.~E., {Rodriguez}, D.~R., {Stevenson}, K.~B., \& {Wakeford}, H.~R.
  2019, Research Notes of the American Astronomical Society, 3, 193

\bibitem[{{National Academies of Sciences, Engineering, and
  Medicine}(2018)}]{NAS-ESS2018}
{National Academies of Sciences, Engineering, and Medicine}. 2018, Exoplanet
  Science Strategy (Washington, DC: The National Academies Press),
  doi:10.17226/25187

\bibitem[{{National Academies of Sciences, Engineering, and
  Medicine}(2019)}]{NAS-AS2019}
---. 2019, An Astrobiology Strategy for the Search for Life in the Universe
  (Washington, DC: The National Academies Press), doi:10.17226/25252

\bibitem[{Seager \& Deming(2009)}]{Seager2009}
Seager, S., \& Deming, D. 2009, The Astrophysical Journal, 703, 1884–1889

\bibitem[{{Seifahrt} {et~al.}(2018){Seifahrt}, {St{\"u}rmer}, {Bean}, \&
  {Schwab}}]{Seifahrt2018}
{Seifahrt}, A., {St{\"u}rmer}, J., {Bean}, J.~L., \& {Schwab}, C. 2018, in
  Society of Photo-Optical Instrumentation Engineers (SPIE) Conference Series,
  Vol. 10702, Proc.~SPIE, 107026D

\bibitem[{Selsis {et~al.}(2011)Selsis, Wordsworth, \& Forget}]{Selsis2011}
Selsis, F., Wordsworth, R.~D., \& Forget, F. 2011, Astronomy \& Astrophysics,
  532, A1

\bibitem[{{Skilling}(2004)}]{Skilling2004}
{Skilling}, J. 2004, in American Institute of Physics Conference Series, Vol.
  735, American Institute of Physics Conference Series, ed. R.~{Fischer},
  R.~{Preuss}, \& U.~V. {Toussaint}, 395--405

\bibitem[{Skilling(2006)}]{Skilling2006}
Skilling, J. 2006, Bayesian analysis, 1, 833

\bibitem[{{Speagle}(2020)}]{Speagle2020}
{Speagle}, J.~S. 2020, MNRAS, 493, 3132

\bibitem[{{Staguhn} {et~al.}(2019){Staguhn}, {Mandell}, {Stevenson}, {Saxena},
  {Kopparapu}, {Fixsen}, {Sharp}, {DiPirro}, {Knez}, {Wolf}, {Sotzen}, {Mandt},
  {Gong}, \& {Villanueva}}]{Staguhn2019}
{Staguhn}, J., {Mandell}, A., {Stevenson}, K., {et~al.} 2019, arXiv e-prints,
  arXiv:1908.02356

\bibitem[{{Stevenson} {et~al.}(2017){Stevenson}, {Line}, {Bean}, {D{\'e}sert},
  {Fortney}, {Showman}, {Kataria}, {Kreidberg}, \& {Feng}}]{Stevenson2017a}
{Stevenson}, K.~B., {Line}, M.~R., {Bean}, J.~L., {et~al.} 2017, aj, 153, 68

\bibitem[{{The LUVOIR Team}(2019)}]{LUVOIR2019}
{The LUVOIR Team}. 2019, {The LUVOIR Mission Concept Study Final Report},
  arXiv:1912.06219

\bibitem[{{Wakeford} {et~al.}(2019){Wakeford}, {Lewis}, {Fowler}, {Bruno},
  {Wilson}, {Moran}, {Valenti}, {Batalha}, {Filippazzo}, {Bourrier},
  {H{\"o}rst}, {Lederer}, \& {de Wit}}]{Wakeford2019}
{Wakeford}, H.~R., {Lewis}, N.~K., {Fowler}, J., {et~al.} 2019, \aj, 157, 11

\bibitem[{{Werner} {et~al.}(2004){Werner}, {Roellig}, {Low}, {Rieke}, {Rieke},
  {Hoffmann}, {Young}, {Houck}, {Brandl}, {Fazio}, {Hora}, {Gehrz}, {Helou},
  {Soifer}, {Stauffer}, {Keene}, {Eisenhardt}, {Gallagher}, {Gautier}, {Irace},
  {Lawrence}, {Simmons}, {Van Cleve}, {Jura}, {Wright}, \&
  {Cruikshank}}]{SPITZER}
{Werner}, M.~W., {Roellig}, T.~L., {Low}, F.~J., {et~al.} 2004, Astrophy. J.
  Suppl. Ser., 154, 1

\end{thebibliography}

%\clearpage
\appendix
%\vspace{-2em}
\subsection{Methods}
%Simulation Setup and Analysis}

%assume perfect knowledge of instrument throughput

We load exoplanet system parameters from \href{https://exo.mast.stsci.edu/}{exo.MAST} and compute missing entries using \href{https://github.com/kevin218/exoMAST\_Obs}{publicly-available code} written for the Exoplanet Atmosphere Observability Table \citep[EAOT,][]{Mullally2019}.
We use Pandexo to generate realistic uncertainties for {\jwst}'s NIRISS/SOSS and NIRSpec/G395H instrument modes.  Our final results use spectra that have been binned to a resolving power of 100.  Tests show that using spectra at the native resolution of the instruments do not impact our results. 
To estimate realistic uncertainties with {\origins} and {\em MIRECLE}, we use custom code originally written by Thomas Greene (personal communication).  This is the same code adopted for the {\em Origins Space Telescope} Mission Concept Study Report \citep{OST2019}.
Since our goal is to assess the magnitude of our parameter uncertainties, we do not add noise to the simulated data.  Doing so would potentially drive the median away from the true value and prevent us from identifying biases in the retrieved results.

\begin{figure}[!t]
    \centering
    \includegraphics[width=0.99\linewidth]{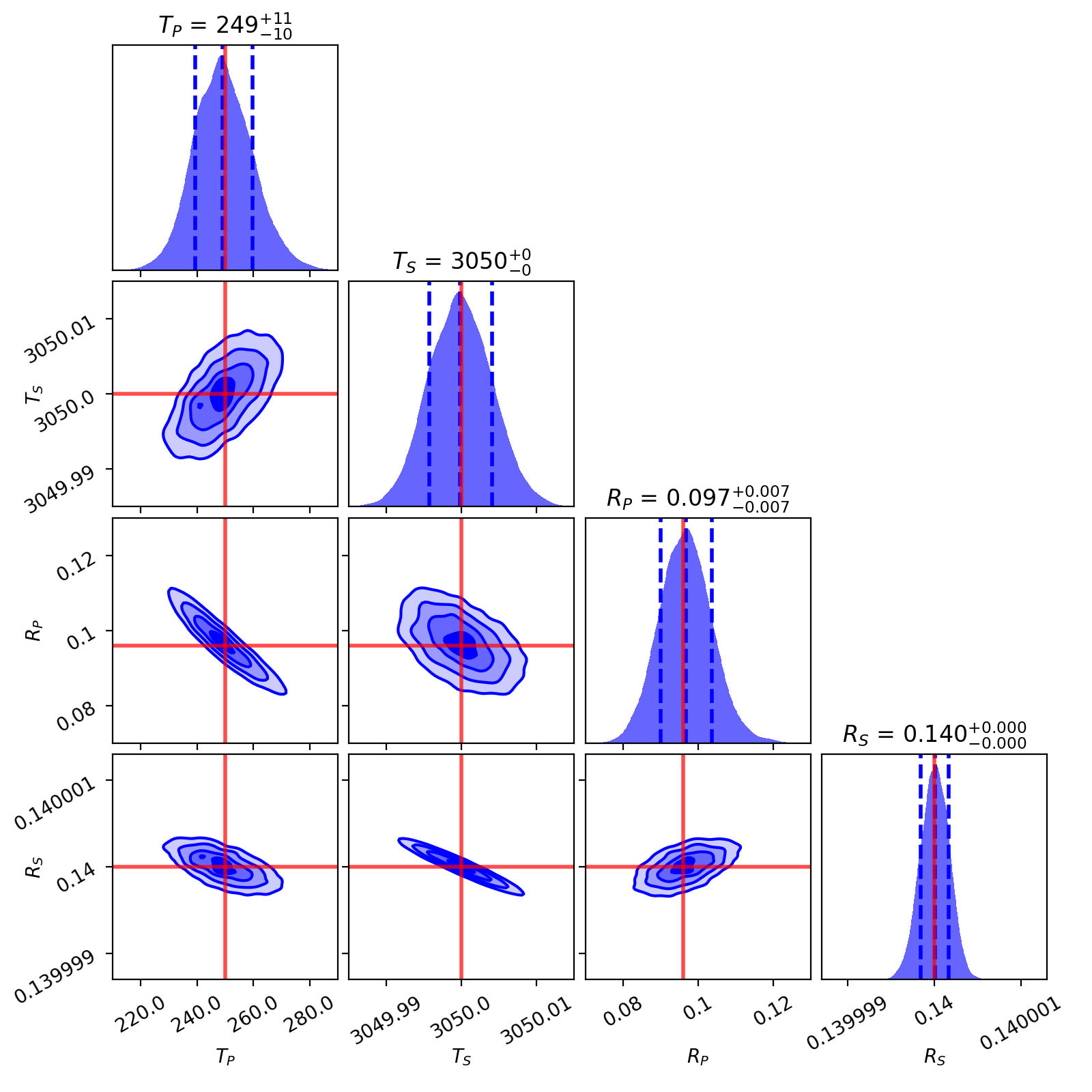}
    %\vspace*{-0.5\baselineskip}
    \caption{\label{fig:corner-ost}{
    {\bf Example correlation pairs plot with 1D marginalized posteriors for a simulation of the Proxima Centauri system using 10 hours of integration time and a 6-meter {\origins}-like telescope.}  In order, the free parameters include planet temperature (K), stellar temperature (K), planet radius (\rjup), and stellar radius (\rsun).  Solid red lines indicate the truth.  Dashed blue lines represent the 16\%, 50\%, and 84\% credibility regions ({\em i.e.}, median with 1$\sigma$ uncertainties).  With precise, broad-wavelength coverage, we are able to minimize the planet's radius-temperature degeneracy.  The stellar parameters in these simulations are more tightly constrained than would be the case if we were to use more realistic model stellar spectra.
    }}
\end{figure}

\begin{figure}[!t]
    \centering
    \includegraphics[width=0.99\linewidth]{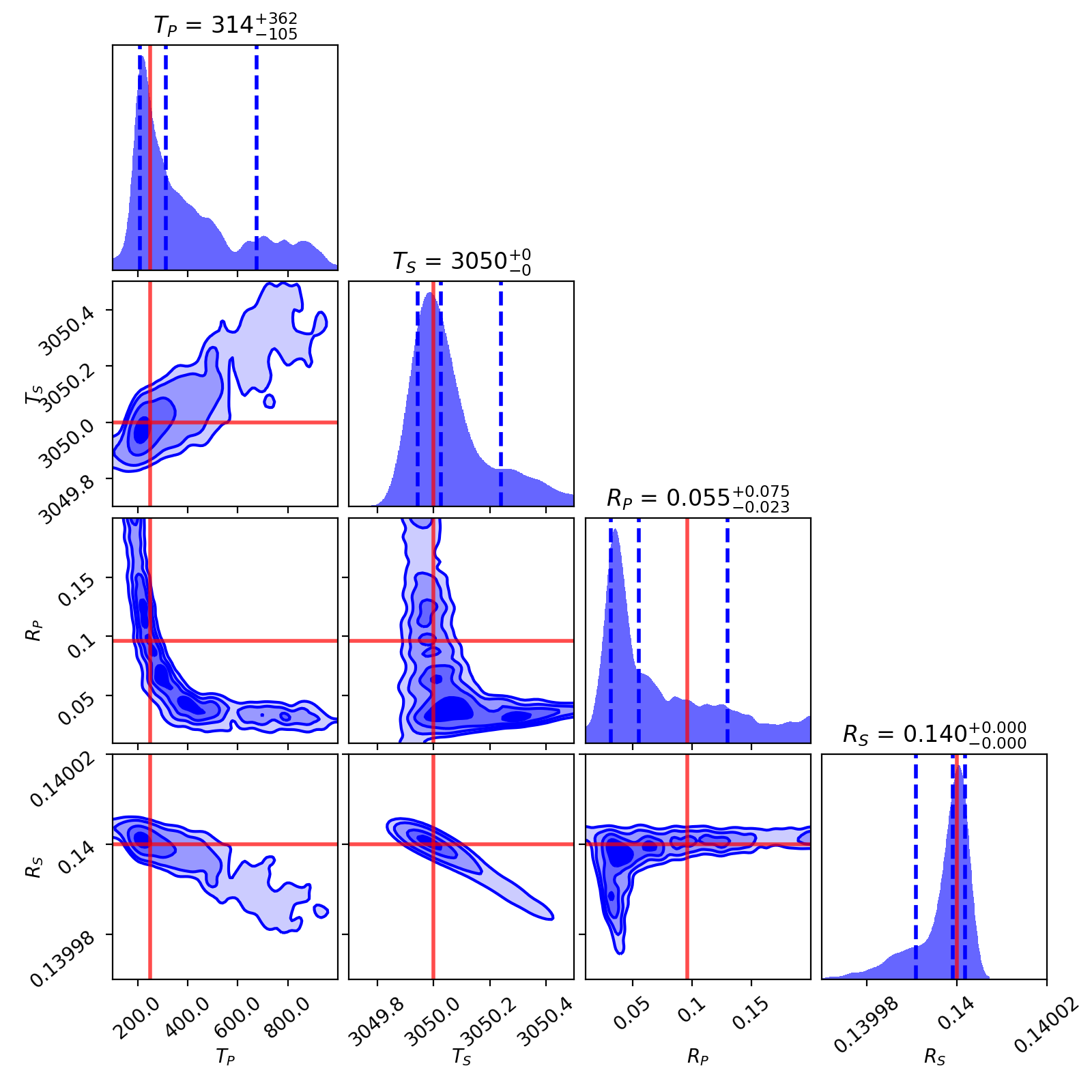}
    %\vspace*{-0.5\baselineskip}
    \caption{\label{fig:corner-jwst}{
    {\bf Same as \Cref{fig:corner-ost}, except using {\jwst}'s MIRI/LRS instrument mode.}  In this case, we are unable to resolve the planet's radius-temperature degeneracy, thus leading to large uncertainties in both parameters.
    }}
\end{figure}

For all of our retrievals, we use a nested-sampling algorithm \citep{Skilling2004,Skilling2006}, as implemented by DYNESTY \citep{Speagle2020}.  For the transiting exoplanet simulations, the log likelihood function uses the reported transit depth to constrain the planet-to-star area ratio.  This has the benefit of providing a tight constraint on the planet radius since the stellar size and temperature are almost always well constrained in our simulations.  We initialize each run using 500 live points spread uniformly across our defined prior region with multiple bounding ellipsoids \citep{Feroz2009}.  Valid prior regions vary for different simulated exoplanet systems, but are always consistent within a related set of simulations.  \Cref{fig:corner-ost} and \Cref{fig:corner-jwst} provide two example correlation pairs plots with 1D marginalized posteriors for {\origins} and {\jwst}, respectively.  These figures demonstrate how parameter degeneracies vary with wavelength coverage.

%Corner plot of Prox Cen with OST

%Corner plot of Prox Cen with JWST

\subsection{Future Work}
%Simulation Limitations

It is important to understand the limits of our blackbody simulations and how to improve upon them.  The simulations presented here have yielded best-case uncertainties due to the simplified (blackbody) representation of our objects and limited number of free parameters (temperature and radius).  Future work will need to incorporate more complex models to address a series of potential obstacles.  These include:

\squishlist
\item Constraining the planet's thermal structure and atmospheric composition using realistic stellar and planetary models;
\item Quantifying the impact of a non-zero, wavelength-dependent planet albedo at various orbital phases;
\item Resolving the potential degeneracy between planet size and orbital inclination;
\item Correcting for achromatic stellar variability; and
\item Characterizing individual planets in multi-planet systems.
\squishend

Upon completion of a more in-depth study, we will be able to better assess the validity of the PIE technique to accurately determine a planet's absolute flux under more realistic conditions.  From there, we can establish the wavelength range, resolving power, and precision necessary to successfully implement the PIE technique with data from future observatories.  These values will likely depend on the stellar type, planet size and temperature, and the scientific goal that one hopes to achieve.  For example, we expect that assessing the habitability of a terrestrial M-dwarf planet will have different requirements than a hot Jupiter orbiting a G dwarf.  Ultimately, it is important to determine which goals can be met using one or more of {\jwst}'s instrument modes.  This theoretical work can then be followed up with analyses using actual {\jwst} data.  If shown to work for hot Jupiters and warm Neptunes with JWST, these benchmarks should enable us to extrapolate the outcome when observing potentially-habitable worlds.

Looking forward, the combined theoretical and observational results can help guide future instrument and mission concept development (such as a 6-meter {\origins} or a smaller MIDEX or Probe-class telescope) to address the findings laid out in the National Academies' 2018 Exoplanet Science Strategy and 2019 Astrobiology Strategy \citep{NAS-ESS2018,NAS-AS2019}.

% \section{Author publication charges} \label{sec:pubcharge}

% The current cost for the different quanta types is available at 
% \url{https://journals.aas.org/article-charges-and-copyright/#author_publication_charges}. 
% Authors may use the ApJL length calculator to get a {\tt rough} estimate of 
% the number of word and float quanta in their manuscript. The calculator 
% is located at \url{https://authortools.aas.org/ApJL/betacountwords.html}.


\end{document}